\documentclass[a4paper,12pt]{article}
\usepackage{amsmath}
\usepackage{graphicx}

\begin{document}

\begin{center}

{\Large \bf Fermion mass hierarchy in a multiple warped braneworld model}\\[20mm]

R. S. Hundi\footnote{E-mail address: tprsh@iacs.res.in} and
Soumitra SenGupta\footnote{E-mail address: tpssg@iacs.res.in}\\
Department of Theoretical Physics,\\
Indian Association for the Cultivation of Science,\\
2A $\&$ 2B Raja S.C. Mullick Road,\\
Kolkata - 700 032, India.\\[20mm]

\end{center}

\begin{abstract}
A geometric understanding of the origin of mass hierarchy among the standard model fermions have been addressed in a multiple
warped spacetime. We show that the requirement of $10^{16}$ order warping between the Planck and the visible branes 
without creating any significant hierarchy among the moduli results into splitting of fermion masses in the standard model. 
Values of the various parameters of the extra dimensional model and the corresponding fermion masses are determined. 
\end{abstract}

\newpage

\section{Introduction}

The standard model (SM) of elementary particles, despite it's spectacular success in explaining Physics upto scale close to
TeV, 
suffers from the well known gauge hierarchy problem, where the Higgs boson
mass receives quantum loop corrections of the order of Planck scale. To restrict the mass of the  Higgs boson within electroweak scale 
we have to fine tune these quantum corrections
at each loop by adjusting the counter terms in the SM. Resolution to this problem has been explored
in the context of supersymmetric and extra dimensional models.

Another problem  encountered within the framework of SM 
is related to the masses of the fermions. The known fermions in the SM have masses ranging from 1 eV to 172 GeV.
The reason why the fermions have different and hierarchical masses is one of the
challenging problems in particle physics. Various approaches to  solve
the fermion mass hierarchy problem have been discussed over a wide range of works \cite{fer-hier}. In this work we propose 
yet another possible solution to
this problem in a multiply warped geometry model which has already been shown to offer a resolution 
of the gauge hierarchy problem in the context of Higgs mass.

Higher dimensional models offer a new explanation for stabilizing the Higgs boson
mass around the electroweak scale. Among these, the model proposed by Randall and
Sundrum (RS) has a warped geometry with one extra spatial dimension \cite{RS}. The extra
spatial dimension
in this model supports two 3-branes and the warping along this dimension
can generate exponential mass scale suppression $\sim 10^{16}$ between these two 3-branes. Hence, one of the 3-branes
can be interpreted as Planck-brane and the other one to be TeV-brane.
By confining the Higgs boson on the TeV-brane, we generate TeV
scale mass for the Higgs field. Subsequently this model was generalized to accommodate more than 
one extra warped dimension and it was qualitatively indicated  that the requirements of the resolution of gauge hierarchy problem as 
well as little hierarchical moduli give rise to the possibility of
resolving the fermion mass hierarchy problem\cite{CS}. Here, we
demonstrate explicitly that in a multiply  warped model, the fermion mass hierarchy problem can indeed be addressed.

In the multiply warped braneworld models \cite{CS}, the number of 3-branes increases
with the number of extra dimensions. Half  of these 3-branes have energy scale clustered around
TeV scale while the scale of the remaining 3-branes would be around Planck scale.
For example, in a 6-dimensional doubly warped model, there will we four
3-branes, out of which two will have scale  around TeV  and the remaining two
would be around the Planck scale. By computing
the fermion Yukawa couplings on different 3-branes clustered around  TeV scale, we show
that the respective masses of fermions will split. Based on this, we argue
that in a multiply warped braneworld model, the fermion mass splitting between these 3-branes can
offer an explanation of the observed fermion mass hierarchy of the SM. More specifically,
we take the 6-dimensional warped model as our case study and
we show that by adjusting the parameters of the model to appropriate  values,
the fermion mass splitting on the two TeV-branes can be generated. By extending this analysis to more than six dimensions, we
show the possibility of explaining all fermion masses of the SM.

The paper is organized as follows. In the next section we give a brief
overview of multiply warped braneworld models. In Sec. 3 we analyze
the fermion mass splitting on two different 3-branes in the 6-dimensional
warped model. We generalize this to higher dimensional multiply warped braneworld models, and
argue that the different fermion masses in the SM can be explained
from the  mass splittings that are generated in these models.
We conclude in Sec. 4.

\section{Multiply warped braneworld model}

In this section we give brief description of six and seven dimensional
warped spacetime models \cite{CS}, which are extended from the 5-dimensional
RS model. Further generalization of these models beyond seven dimensions can
be worked out in a similar fashion \cite{CS}.

The six dimensional warped model has the usual four spacetime
and two extra spatial dimensions. The two extra spatial dimensions
are successively compactified on $S_1/Z_2$ orbifold. The manifold of the six dimensional
spacetime is $[M^{1,3}\times S^1/Z_2]\times S^1/Z_2$. The metric ansatz consistent
with this manifold is
\begin{equation}
ds^2 = b^2(z)[a^2(y)\eta_{\mu\nu}dx^\mu dx^\nu +R_y^2dy^2]+r_z^2dz^2,
\label{E:metric}
\end{equation}
where $x^\mu$, $\mu=0,\cdots,3$, are the four non-compact spacetime coordinates,
$y$ and $z$ are the angular coordinates of the two extra dimensions with
the moduli $R_y$ and $r_z$, respectively. $a(y)$ and $b(z)$ are some functions
of the extra coordinates. Here, $\eta_{\mu\nu} = {\rm diag}(-1,1,1,1)$.
Since the $y$ and $z$ coordinates are orbifolded on $S^1/Z_2$, with the orbifold
fixed points $y=0,\pi$ and $z=0,\pi$, there are four 4-branes at these fixed points.
Intersection of any of the two 4-branes give one 3-brane. Hence, in this
model there would be four different 3-branes located at $(y,z)=(\pi,\pi)$,
$(\pi,0)$, $(0,\pi)$ and $(0,0)$. We take one of these 3-branes
as our Tev scale universe.
The total bulk-brane action of this model has a form \cite{CS}
\begin{eqnarray}
S &=& S_6+S_5,\quad
%\nonumber \\
S_6 = \int d^4xdydz\sqrt{-g_6}(R_6-\Lambda),
\nonumber \\
S_5 &=& \int d^4xdydz[V_1\delta(y)+V_2\delta(y-\pi)]
     +\int d^4xdydz[V_3\delta(z)+V_4\delta(z-\pi)]
\end{eqnarray}
Here, $V_{1,2}$ and $V_{3,4}$ are brane tensions of the branes located at
$y=0,\pi$ and $z=0,\pi$, respectively. $\Lambda$ is the cosmological
constant in 6-dimensions. \\
After solving Einstein's
equations with orbifolded boundary conditions the solution to the warp functions
of the metric of Eq. (\ref{E:metric}) has a form \cite{CS}
\begin{eqnarray}
a(y) &=& \exp(-c|y|), \quad b(z) = \frac{\cosh(kz)}{\cosh(k\pi)},
\nonumber \\
c&\equiv &\frac{R_yk}{r_z\cosh(k\pi)},\quad k\equiv r_z\sqrt{
\frac{-\Lambda}{10M_P^4}}.
\label{E:sol}
\end{eqnarray}
Here, $M_P$ is the Planck scale.

The functional form of $a(y)$ in Eq. (\ref{E:sol}) is similar
to the exponential warping in the 5-dimensional RS model \cite{RS}. In addition
the function $b(z)$ gives warping along $z$ direction. It can be
noticed that for the 3-brane located at $y = 0, z = \pi$ we have no warping and hence it is called Planck brane while
for  the 3-brane located at $y = \pi, z = 0$ the warping is maximum and is called standard model or TeV brane.
However, from the relation for $c$ in Eq. (\ref{E:sol}),
in order not to have large hierarchy between $R_y$ and $r_z$, one of 
$c$ or $k$ must be large and other is small. This indicates large warping in one direction and small along the other.
It is now easy to show that when 
Eq. (\ref{E:sol}) is considered together with the maximum warping condition at the 
visible brane i.e. $a(\pi) b(0) = 10^{-16}$ then the minimum hierarchy between the two moduli
can be achieved for $c\sim 11$ and $k\sim 0.4$. Now, with this choice
of values for $c$ and $k$, the 3-branes located at $(y,z)=(\pi,0)$ and
$(\pi,\pi)$ get exponential suppression $\sim 10^{-16}$ with respect to the  3-branes located
at $(y,z)=(0,0)$ and $(0,\pi)$.
However due to small warping along z-direction, the mass scales on the 3-branes at $(y,z)=(0,0)$
and $(0,\pi)$ would of the order of Planck scale while the mass scales on the
3-branes at $(y,z)=(\pi,0)$ and $(\pi,\pi)$ would be
suppressed close to TeV scale. So in the six dimensional warped model,
the requirement of absence of any large hierarchy between the two moduli 
along with both the moduli implies that the scales of two 3-branes are clustered around TeV scale and 
that of the other two are at around Planck scale. The moduli of such six dimensional models can be stabilized 
to their desired values following the mechanism suggested by Goldberger and Wise 
where a bulk scalar field is introduced as a stabilizing field with appropriate vacuum expectation values at the boundaries \cite{GW}

We now describe the seven dimensional model which has three extra
spatial dimensions. Each of the extra dimension is orbifolded by $Z_2$
symmetry. The manifold under consideration of this model is
$[\{M^{1,3}\times S^1/Z_2\}\times S^1/Z_2]\times S^1/Z_2$. Denoting
the three extra dimensions by angular coordinates $y$, $z$ and $w$,
the bulk-brane action in this model can be written as
\begin{eqnarray}
S&=&S_7+S_6,\quad S_7 = \int d^4xdydzdw\sqrt{-g_7}(R_7-\Lambda_7),
\nonumber \\
S_6 &=& \int d^4xdydzdw[V_1\delta(w)+V_2\delta(w-\pi)]
     +\int d^4xdydzdw[V_3\delta(z)+V_4\delta(z-\pi)]
\nonumber \\
&& +\int d^4xdydzdw[V_5\delta(y)+V_6\delta(y-\pi)].
\end{eqnarray}
Here, $\Lambda_7$ is cosmological constant in 7-dimensions. $V_{1,\cdots,6}$
are brane tensions of 5-branes located at $w=0,\pi$, $z=0,\pi$ and $y=0,\pi$,
respectively. The metric solution of the above action which satisfies
Einstein's equations and satisfies the orbifolded boundary condition is \cite{CS}
\begin{eqnarray}
ds^2&=&\frac{\cosh^2(lw)}{\cosh^2(l\pi)}\left\{\frac{\cosh^2(kz)}{\cosh^2(k\pi)}
\left[\exp(-2cy)\eta_{\mu\nu}dx^\mu dx^\nu +R_y^2dy^2\right]+r_z^2dz^2\right\}
+R_w^2dw^2
\nonumber \\
l^2&\equiv &\frac{\Lambda_7R_w^2}{15},\quad k\equiv\frac{lr_z}{R_w\cosh(l\pi)},
\nonumber \\
c&\equiv &\frac{lR_y}{R_w\cosh(k\pi)\cosh(l\pi)} = \frac{kR_y}{r_z\cosh(k\pi)}.
\end{eqnarray}
Here, $R_w$ is the moduli corresponding to the $w$ direction. Looking at the above
solution of metric, we have exponential suppression along the $y$ direction,
whereas, the suppression along $z$ and $w$ are given  by cosine-hyperbolic
functions. Once again a  consistent set of values for the dimensionless quantities
would be: $c\sim 10$, $k\sim 0.1$ and $l\sim 0.1$, so that there will not
be large hierarchies among the moduli $R_y$, $r_z$ and $R_w$.

In the seven dimensional model, the 5-branes located at $w=0,\pi$, $z=0,\pi$
and $y=0,\pi$ intersect and give twelve different 4-branes which lie at
$(z,w)=(0,0),(0,\pi),(\pi,0),(\pi,\pi)$, $(z,y)=(0,0),(0,\pi),(\pi,0),(\pi,\pi)$
and $(y,w)=(0,0),(0,\pi),(\pi,0),(\pi,\pi)$. Intersection of 4-branes results
in 3-branes, and in this model there are eight different possibilities,
which lie at $(y,z,w)=(0,0,0),(0,0,\pi),(0,\pi,0),(\pi,0,0),(\pi,\pi,0),(\pi,0,\pi),
(0,\pi,\pi)$ and $(\pi,\pi,\pi)$. For the consistent set of values of $c$, $k$ and $l$,
which are described in the previous paragraph, the 3-branes located at
$(0,0,0),(0,0,\pi),(0,\pi,0)$ and $(0,\pi,\pi)$ have no 
suppression, hence, these 3-branes can be identified as Planck scale branes.
Whereas, the 3-branes located at $(\pi,0,0),(\pi,0,\pi),(\pi,\pi,0)$ and
$(\pi,\pi,\pi)$ have exponential suppression $\sim 10^{16}$ to  TeV scale masses,
and hence they may be identified as TeV scale branes. So in the seven
dimensional model, four 3-branes are clustered around TeV scale and the remaining
four are clustered around Planck scale.

The six and seven dimensional braneworld models, which are described here,
can easily be generalized to higher dimensions. A notable feature
in all these models is that half of the total number of 3-branes would be clustered
around TeV scale and the remaining 3-branes would be clustered around
Planck scale. As explained in Sec. 1, we exploit this feature and try
to see how much splitting between fermion masses can be obtained geometrically. 
In the next section we describe the calculation
of fermion mass splitting in the six dimensional model. We can generalize
this calculation to higher dimensions, and we argue that in a multiply
braneworld model the hierarchical mass pattern of SM fermions can
be explained.

\section{Fermion mass splitting in a multiply warped braneworld model}

As explained previously, we take 6-dimensional warped model as our case
study and calculate fermion mass splitting between two different
3-branes. A viable 
way of generating fermion masses is through Yukawa couplings. We generalize
the approach taken in \cite{GN}, for generating fermion masses
in multiply braneworld models.
We assume that both the Higgs boson and left-handed doublet quark fields
of SM are confined to the usual 4-dimensional world. We allow the
right-handed fields of the SM to
be in the bulk of 6-dimensions. Consider the Yukawa couplings for
left-handed doublets with its right-handed counter-parts. Since
right-handed fermions are allowed in the bulk, we can estimate
the gauge invariant Yukawa couplings on both the 3-branes at
$(y,z)=(\pi,0),(\pi,\pi)$, which are given below.
\begin{eqnarray}
S_{\rm vis} &=&\int d^4x\sqrt{-g_{\rm vis}}\hat{Y}_6\bar{L}_{0L}(x)H_0(x)\psi_R(x,\pi,0) +{\rm h.c.},
\nonumber \\
S_{\rm vis^\prime}&=&\int d^4x\sqrt{-g_{\rm vis^\prime}}\hat{Y}^\prime_6\bar{L}_{0L}(x)H_0(x)
\psi_R(x,\pi,\pi) +{\rm h.c.}.
\label{E:feract}
\end{eqnarray}
Here, $\hat{Y}_6,\hat{Y}^\prime_6$ are Yukawa couplings in 6-dimensions in the two branes. Note
that their dimensions are $M^{-1}$, where $M\sim M_P$. $L_{0L}$ indicates
left-handed SU(2) doublet of either quark or lepton field of the SM, and $\psi_R$ is
its right-handed counter-part. $H_0$ is the Higgs doublet. $g_{\rm vis} =
{\rm det}(g^6_{\mu\nu}(\pi,0))$ and $g_{\rm vis^\prime}={\rm det}(g^6_{\mu\nu}(\pi,\pi))$,
where $g^6_{M,N}$ is the metric function defined in Eq. (\ref{E:metric}).
Since the fermions are 8-component spinors in 6-dimensions, we have to
choose the representation for Higgs field $H_0$ accordingly in the above
equation. Also since $\psi_R$ is a bulk field, it decomposes into Kaluza-Klein
(KK) fields in the 4-dimensional world.
After the Higgs acquires vacuum expectation value, the Yukawa
couplings in above equation give mixing masses between the left-handed
fermion and the KK tower of right-handed fields. The zero mode of the
KK tower is identified as the SM field. Nevertheless, to be consistent
we have to include all possible mixings with the KK fermions, and after
diagonalizing it we can identify the lowest eigenvalue to be the SM
fermion mass. In fact, this is the procedure employed in \cite{GN},
and here we generalize this to 6-dimensions. For this we have to
study the KK excitations of a bulk fermion field, which is done in
\cite{DHS} for a  6-dimensional model. However, the nonzero KK masses
and their wave functions are not computed in \cite{DHS}. In the next
subsection we calculate them which will be necessary for the computation
of SM fermion masses.

\subsection{KK fermions}

The kinetic part of the action for a bulk fermion field $\Psi$
in 6-dimensions is \cite{GN,DHR,CHNOY}
\begin{equation}
S_f = \int d^4xdydz\sqrt{-G}E^A_a\left[\frac{i}{2}\left(\bar{\Psi}\Gamma^a
\partial_A\Psi - \partial_A\bar{\Psi}\Gamma^a\Psi\right)+\frac{\omega_{bcA}}{8}
\bar{\Psi}\{\Gamma^a,\sigma^{bc}\}\Psi\right]
\label{E:Sf}
\end{equation}
Here, the capital letter $A$ denotes index in the curved space
and the lower case letters $a,b,c$ denote indices in the
tangent space.
$\omega_{bcA}$ is spin connection and $E^A_a$ is inverse vielbein. Here,
we do not write the bare mass of bulk fermion 
since the SM fermions acquire mass through the Higgs mechanism via Eq. (\ref{E:feract}).
The spin connection term in the
above equation give zero contribution due to the diagonal form of the metric in
Eq. (\ref{E:metric}).
The order of the Dirac matrices $\Gamma^a$ in 6-dimensions would be 8$\times$8, and
they can be taken as \cite{BDP}:
\begin{equation}
\Gamma^\mu = \gamma^\mu\otimes\sigma^0,\quad \Gamma^4=i\gamma_5\otimes\sigma^1,
\quad \Gamma^5=i\gamma_5\otimes\sigma^2
\end{equation}
Here, $\gamma^\mu$ are the Dirac matrices in 4-dimensions and
$\gamma_5=i\gamma^0\gamma^1\gamma^2\gamma^3$. $\sigma^i$, $i=1,2,3$, are the
Pauli matrices and $\sigma^0$ is the 2$\times$2 unit matrix.
The chirality in 6-dimensions is defined by the matrix $\bar{\Gamma}=\Gamma^0
\Gamma^1\Gamma^2\Gamma^3\Gamma^4\Gamma^5$ such that $\bar{\Gamma}\Psi_\pm = \pm\Psi_\pm$.
The chiral fermions in 6-dimensions have both left- and right-handed chirality
of 4-dimensions, which can be projected by the operators $P_{L,R} =
(1\mp i\Gamma^0\Gamma^1\Gamma^2\Gamma^3)/2$. Since, fermions have
chirality in six dimensions, we can write $\psi=\psi_++\psi_-$. Each
field carrying either plus or minus chirality can further be decomposed
into left- ($\psi_{\pm L}$) and right-chiral ($\psi_{\pm R}$) fields in four dimensions.
These fields can be expressed in KK expansion as \cite{DHS}
\begin{eqnarray}
\Psi_{+L,-R}(x^\mu,y,z)&=&\frac{1}{\sqrt{R_yr_z}}\sum_{j,k}\psi^{(j,k)}_{+L,-R}(x^\mu)
f^{(j,k)}_{+L,-R}(y,z)\otimes\left(\begin{array}{c}1\\0\end{array}\right),
\nonumber \\
\Psi_{-L,+R}(x^\mu,y,z)&=&\frac{1}{\sqrt{R_yr_z}}\sum_{j,k}\psi^{(j,k)}_{-L,+R}(x^\mu)
f^{(j,k)}_{-L,+R}(y,z)\otimes\left(\begin{array}{c}0\\1\end{array}\right).
\label{E:KKfer}
\end{eqnarray}
Here, fields of the form $\psi^{(j,k)}(x^\mu)$ are the KK fields in 4-dimensions and
the KK wave functions should satisfy the following normalization condition and
eigenvalue equations \cite{DHS}.
\begin{equation}
\int dydz b^4(z)a^3(y)\left(f^{(j,k)}_{+R,+L,-R,-L}(y,z)\right)^*
f^{(j^\prime,k^\prime)}_{+R,+L,-R,-L}(y,z) =
\delta^{j,j^\prime}\delta^{k,k^\prime},
\label{E:fnorm}
\end{equation}
\begin{eqnarray}
(i{\cal D}_y+{\cal D}_z)f^{(j,k)}_{+R}(y,z)&=&-M_{j,k}f^{(j,k)}_{+L}(y,z),
\nonumber \\
(-i{\cal D}_y+{\cal D}_z)f^{(j,k)}_{+L}(y,z)&=&-M_{j,k}f^{(j,k)}_{+R}(y,z),
\nonumber \\
(i{\cal D}_y+{\cal D}_z)f^{(j,k)}_{-L}(y,z)&=&M_{j,k}f^{(j,k)}_{-R}(y,z),
\nonumber \\
(-i{\cal D}_y+{\cal D}_z)f^{(j,k)}_{-R}(y,z)&=&M_{j,k}f^{(j,k)}_{-L}(y,z),
\label{E:feigen}
\end{eqnarray}
where the differential operators: ${\cal D}_y=\frac{i}{2R_y}(4\partial_ya
+2a\partial_y)$, ${\cal D}_z=\frac{i}{2r_z}a(5\partial_zb
+2b\partial_z)$
and  $M_{j,k}$ is the mass of the KK fermions $\psi^{(j,k)}$.
In \cite{DHS} the KK wave function for zero mode has been solved, which
is identified with $M_{0,0}=0$. By writing $f_{+R}^{(0,0)}(y,z)=f_y(y)f_z(z)$,
we get the solution for $f_y$ and $f_z$ as
\begin{equation}
f_y(y)=\exp\left(\frac{1}{2}(c_1+4c)y\right),\quad
f_z(z)=\frac{\exp\left(\frac{-ic_1r_z}{kR_y}\tan^{-1}(\tanh(kz/2))\cosh(k\pi)\right)}
{\cosh^{5/2}(kz)}
\label{E:SMwf}
\end{equation}
Here, $c_1$ is a separation constant which is defined in \cite{DHS}, and
can be worked out from the normalization of wave function
$f_{+R}^{(0,0)}(y,z)$, using Eq. (\ref{E:fnorm}). The nonzero KK masses
and their wave functions are necessary to calculate the mixing masses
of Eq. (\ref{E:feract}), which will be worked out below.

The eigenvalue equations in Eq. (\ref{E:feigen}) can be transformed into
second order differential equations as
\begin{eqnarray}
(-i{\cal D}_y+{\cal D}_z)(i{\cal D}_y+{\cal D}_z)f_{+R,-L}^{(j,k)} &=&
M_{j,k}^2f_{+R,-L}^{(j,k)},
\nonumber \\
(i{\cal D}_y+{\cal D}_z)(-i{\cal D}_y+{\cal D}_z)f_{-R,+L}^{(j,k)} &=&
M_{j,k}^2f_{-R,+L}^{(j,k)}.
\end{eqnarray}
Here we will show how the nonzero eigenvalues $M_{j,k}$ can be found
for the wave function $f_{+R}^{(j,k)}$. The wave functions of other
chiral modes can be analogously worked out. By writing $f_{+R}^{(j,k)}
=\chi^{(j,k)}(y)\xi^{(j,k)}(z)$, the above second order differential
equation can be recast into
\begin{eqnarray}
&&\frac{1}{(2R_y)^2}\frac{[24c^2-20c\partial_y+4\partial_y^2]\chi^{(j,k)}}{\chi^{(j,k)}}
+\frac{M_{j,k}^2}{a^2}
\nonumber \\
&&+\frac{1}{(2r_z)^2}\frac{(5\partial_zb+2b\partial_z)^2\xi^{(j,k)}}{\xi^{(j,k)}}
+\frac{2ic}{2R_y2r_z}\frac{(5\partial_zb+2b\partial_z)\xi^{(j,k)}}{\xi^{(j,k)}} = 0,
\end{eqnarray}
where the first and second lines of the above equation depends on functions
of $y$ and $z$, respectively. Hence, we can choose
\begin{eqnarray}
\frac{1}{(2R_y)^2}\frac{[24c^2-20c\partial_y+4\partial_y^2]\chi^{(j,k)}}{\chi^{(j,k)}}
+\frac{M_{j,k}^2}{a^2} &=& m_\chi^2,
\nonumber \\
\frac{1}{(2r_z)^2}\frac{(5\partial_zb+2b\partial_z)^2\xi^{(j,k)}}{\xi^{(j,k)}}
+\frac{2ic}{2R_y2r_z}\frac{(5\partial_zb+2b\partial_z)\xi^{(j,k)}}{\xi^{(j,k)}} &=& -m_\chi^2,
\label{E:sepeq}
\end{eqnarray}
where $m_\chi^2$ is a separation constant of mass-square dimensions.
By making the transformations
\begin{equation}
\chi^{(j,k)}=e^{\frac{5}{2}cy}\tilde{\chi}^{(j,k)},\quad
Z_p=\frac{M_{j,k}}{k^\prime}\cosh(k\pi)e^{cy},\quad k^\prime=k/r_z,
\end{equation}
the first of Eq. (\ref{E:sepeq}) can be put into a Bessel's equation as
\begin{equation}
Z_p^2\frac{\partial^2\tilde{\chi}^{(j,k)}}{\partial Z_p^2}+Z_p
\frac{\partial\tilde{\chi}^{(j,k)}}{\partial Z_p}+(Z_p^2-\nu^2)\tilde{\chi}^{(j,k)} = 0,
\end{equation}
where
\begin{equation}
\nu^2 = \frac{25}{4}+\left(m_\chi^2-\frac{6c^2}{R_y^2}\right)\frac{\cosh^2(k\pi)}{k^{\prime^2}}.
\end{equation}
Now, the solution to the wave function $\chi^{(j,k)}(y)$ can be written as
\begin{equation}
\chi^{(j,k)}(y)=\frac{e^{\frac{5}{2}cy}}{N}[J_\nu(Z_p)+bY_\nu(Z_p)],
\end{equation}
where $N$ and $b$ are some constants. $N$ can be determined from the
normalization condition of Eq. (\ref{E:fnorm}). $b$ and the KK masses
$M_{j,k}$ can be determined by applying the following boundary
conditions $\partial_y\chi^{(j,k)}(y)|_{y=0,\pi} = 0$.
The boundary condition at $y=0$ gives us that the constant $b$ is negligibly
small and the condition at $y=\pi$ gives the following relation
\begin{equation}
5J_\nu(\tilde{y})+\tilde{y}[J_{\nu -1}(\tilde{y}) - J_{\nu +1}(\tilde{y})] = 0,
\label{E:eigy}
\end{equation}
where $\tilde{y}=\frac{M_{j,k}}{k^\prime}\cosh(k\pi)e^{c\pi}$. The above
equation determines the nonzero mass eigenvalues $M_{j,k}$ of the KK modes
for a specific value of $\nu$. The index $\nu$ depends on the separation
constant $m_\chi^2$. This value can be determined by solving the
equation along the $z$, which is the second of Eq. (\ref{E:sepeq}).

The second of Eq. (\ref{E:sepeq}) can be expressed as
\begin{equation}
(5\partial_zb+2b\partial_z)[(5\partial_zb+2b\partial_z)\xi^{(j,k)}+k_1\xi^{(j,k)}]
=k_2\xi^{(j,k)}
\end{equation}
where $k_1=2icr_z/R_y$ and $k_2=-4r_z^2m_\chi^2$. The solution
to the above equation can be found from the linear equation
$(5\partial_zb+2b\partial_z)\xi^{(j,k)}=k_3\xi^{(j,k)}$, where
$k_3$ satisfies the quadratic equation $k_3(k_3+k_1)=k_2$. After
doing this, the general solution to the second of Eq. (\ref{E:sepeq})
can be written as
\begin{eqnarray}
\xi^{(j,k)}(z)&=&a_1\xi^+(z)+a_2\xi^-(z),
\nonumber \\
\xi^\pm(z)&=&\frac{\exp(k_3^\pm\tan^{-1}[\tanh(\frac{kz}{2})]\cosh(k\pi)/k)}{\cosh^{5/2}(kz)},
\nonumber \\
k_3^\pm&=&\frac{icr_z}{R_y}[-1\pm\sqrt{1+\frac{4R_y^2m_\chi^2}{c^2}}],
\end{eqnarray}
where $a_{1,2}$ are constants, and one of them can be determined by the 
boundary condition $\partial_z\xi^{(j,k)}(z)|_{z=0,\pi} = 0$. The condition
at $z=0$ gives us $a_1=-a_2k_3^-/k_3^+$. And the condition at $z=\pi$ gives
the following relation
\begin{equation}
k_3^-(k_3^+-5k\tanh(k\pi))\xi^+(\pi)-k_3^+(k_3^--5k\tanh(k\pi))\xi^-(\pi)=0.
\label{E:eigz}
\end{equation}
The undetermined constant $a_2$ can be found from the normalization
condition of Eq. (\ref{E:fnorm}). The left-hand side of Eq. (\ref{E:eigz})
has a sinusoidal behavior, and zeros of this function determine
the nonzero values of $m_\chi^2$, which should be put in Eq. (\ref{E:eigy})
to find the nonzero KK masses $M_{j,k}$ of fermions.

\subsection{Fermion mass splitting}

After calculating the KK masses and their wave functions we now return
to Eq. (\ref{E:feract}), and calculate the SM fermion masses
on both the 3-branes at $(y,z)=(\pi,0),(\pi,\pi)$. To make the kinetic term canonical 
we need to do rescaling of the SM doublet
$L_{0L}$ and the Higgs field $H_0$, since we have confined them
in the 4-dimensional world. Imagine that our universe is identified
as $(y,z)=(\pi,0)$, then the kinetic energy terms of a fermion $\psi_0$
and scalar $\phi_0$ on this brane are
\begin{equation}
S=\int d^4x\sqrt{-g_{\rm vis}}\left[g_{\rm vis}^{\mu\nu}\partial\phi_0^*\partial_\nu\phi_0+
E^\mu_a(\pi,0)\bar{\psi}_0\Gamma^a\partial_\mu\psi_0+\cdots\right]
\end{equation}
The inverse vielbein is given by $E^\mu_a(\pi,0)=\delta^\mu_ae^{c\pi}\cosh(k\pi)$
and we have $g_{\rm vis}^{\mu\nu}=e^{2c\pi}\cosh^2(k\pi)\eta^{\mu\nu}$.
Substituting them in the above equation, we get
\begin{equation}
S=\int d^4x\left[\frac{e^{-2c\pi}}{\cosh^2(k\pi)}\eta^{\mu\nu}\partial\phi_0^*\partial_\nu\phi_0
+\frac{e^{-3c\pi}}{\cosh^3(k\pi)}\bar{\psi}_0\Gamma^\mu\partial_\mu\psi_0+\cdots\right]
\end{equation}
Hence, to get the canonical kinetic terms for scalar and fermion
fields in 4-dimension, we have to do the following rescalings
\begin{equation}
\phi_0= e^{c\pi}\cosh(k\pi)\phi,\quad\psi_0= e^{\frac{3}{2}c\pi}\cosh^{3/2}(k\pi)\psi,
\end{equation}
where $\phi$ and $\psi$ are renormalized scalar and fermion fields.
Recall Sec. 2 where  we pointed out that $c\sim 10$ and $k\sim 0.1$ give
a consistent set of values. For this set of values $e^{c\pi}\cosh(k\pi)\sim 10^{16}$,
which naturally gives $\langle\phi\rangle\sim$ 100 GeV. This is in similar
lines of what it is shown in the 5-dimensional RS model \cite{GN}.

The SU(2) doublet fermion of the SM can be put in either of the $+$ or the $-$ chirality
of the 6-dimensional world. Without any loss of generality, we choose it to
be in the $-$ chirality of 6-dimensional spinor. Then, after substituting
the KK wave function expansion for $\Psi_R$, Eq. (\ref{E:KKfer}), and also
doing the rescaling of the 4-dimensional fields $L_{0L}$ and $H_0$, the
first of Eq. (\ref{E:feract}) turns out to be
\begin{equation}
S_{\rm vis}=\sum_{j,k}\int d^4x\frac{e^{-\frac{3}{2}c\pi}}{\cosh^{3/2}(k\pi)}
f_{+R}^{(j,k)}(\pi,0)Y_6\bar{L}_LH\psi_{+R}^{(j,k)}.
\end{equation}
Here, $Y_6=\hat{Y}_6/\sqrt{R_yr_z}$. $Y_6$ is the effective 4-dimensional
Yukawa coupling which is an ${\cal O}(1)$ quantity. After the Higgs field
acquires vacuum expectation value (vev), the left-handed SM fermion gets mixing
masses with all the right-handed KK fields, which can be written as
\begin{equation}
[m_D]_{ij}=\frac{e^{-\frac{3}{2}c\pi}}{\cosh^{3/2}(k\pi)}f_{+R}^{(j,k)}(\pi,0)Y_6v,
\label{E:effDir}
\end{equation}
where $v$ is the vev of Higgs field. In the previous subsection, we have seen
that the fields $\psi_{+R}^{(j,k)}$ have Dirac masses with their left-handed
counter parts. Hence, in the basis $\Psi_L=(\psi_L,\psi_{+L}^{(1,0)},\psi_{+L}^{(0,1)},
\psi_{+L}^{(1,1)},\cdots)$,
$\Psi_R=(\psi_R,\psi_{+R}^{(1,0)},\psi_{+R}^{(0,1)},\psi_{+R}^{(1,1)},\cdots)$,
where $\psi_L$ is a component of the
SM doublet $L_L$ and $\psi_R=\psi^{(0,0)}_{+R}$, we get the following form
for mixing matrix $\bar{\Psi}_L{\cal M}\Psi_R+{\rm h.c.}$, 
\begin{equation}
{\cal M}=\left(\begin{array}{ccccc}
[m_D]_{0,0} & [m_D]_{1,0} & [m_D]_{0,1} & [m_D]_{1,1} & \cdots \\
0 & M_{1,0} & 0 & 0 & \cdots \\
0 & 0 & M_{0,1} & 0 & \cdots \\
0 & 0 & 0 & M_{1,1} & \cdots \\
\vdots & \vdots & \vdots & \vdots & \ddots
\end{array}\right)
\end{equation}
We have numerically computed the nonzero KK masses $M_{i,j}$, whose values are coming
in the multi-TeV scale. From the relations given in previous subsection,
the KK masses depend on $r_z$ apart from $k$ and $c$. For $r^{-1}_z\sim 10^{18}$ GeV,
the lowest KK mass has been found to be around 6 TeV. Here we have fixed $k=0.7$
and $c=11.24$ where the justification of these values will be given later
(see Table \ref{T:1}). Some of the modes like $M_{0,1},M_{0,2}$, etc. may
decouple away since their masses and wave functions are found out to be close to zero.
We have also found that the off-diagonal values of $[m_D]_{i,j}$
are coming in the electroweak scale. As a result due to the high values of the KK masses,
after diagonalizing the above matrix, we can see that the zeroth mode mass
$[m_D]_{0,0}$ is negligibly affected. Hence, the SM
fermion mass is approximately the same as the zeroth mode mass $[m_D]_{0,0}$.
The same kind of feature can be seen if we repeat this calculation on the
other 3-brane at $(y,z)=(\pi,\pi)$, i.e. on the second term of
Eq. (\ref{E:feract}). Because of this, to calculate the fermion mass splitting
between the two 3-branes, $(y,z)=(\pi,0),(\pi,\pi)$, we work out the
corresponding values of the zeroth mode.

From Eq. (\ref{E:effDir}) we can notice that the following factor
\begin{equation}
F=\frac{e^{-\frac{3}{2}c\pi}}{\cosh^{3/2}(k\pi)}f_{+R}^{(0,0)}(\pi,0)
=\frac{e^{\frac{1}{2}(c_1+c)\pi}}{\cosh^{3/2}(k\pi)}
\end{equation}
determines the zeroth mode mass on the 3-brane at $(y,z)=(\pi,0)$ where
$Y_6v\sim$ 100 GeV. In the above equation we have used
$f_{+R}^{(0,0)}(y,z)=f_y(y)f_z(z)$ and also Eq. (\ref{E:SMwf}).
The corresponding factor on the 3-brane at $(y,z)=(\pi,\pi)$ would be
\begin{equation}
F^\prime=e^{-\frac{3}{2}c\pi}\cosh^{5/2}(k\pi)|f_{+R}^{(0,0)}(\pi,\pi)|
=e^{\frac{1}{2}(c_1+c)\pi}.
\end{equation}
The ratio of masses confined on 3-branes at $(y,z)=(\pi,0),(\pi,\pi)$,
is
\begin{equation}
R = \frac{F}{F^\prime} = \frac{1}{\cosh^{3/2}(k\pi)}
\end{equation}
This ratio $R$ gives us the splitting between the fermions on these
two 3-branes. The mass term of a fermion confined on
$(y,z)=(\pi,0)$ brane and the mass term of a fermion confined on the other
TeV 3-brane at $(y,z)=(\pi,\pi)$ will have a relative suppression by
a factor given by $R$. In Table \ref{T:1}, we give possible values of $R$
in the 6-dimensional model. Here we have chosen $c$ so that
$e^{c\pi}\cosh(k\pi)=10^{16}$, which yields  the desired
suppression from Planck to TeV scale in the 6-dimensional model \cite{DHS}.
\begin{table}[!h]
\begin{center}
\begin{tabular}{||c|c|c|c||} \hline
$k$ & $c$ & $\frac{R_y}{r_z}$ & $R$ \\ \hline
0.7 & 11.24 & 73.3 & $\frac{1}{9.75}$ \\
1.2 & 10.75 & 194.34 & $\frac{1}{101.1}$ \\
1.7 & 10.24 & 628.9 & $\frac{1}{1065.7}$ \\
2.2 & 9.75 & 2223.7 & $\frac{1}{11243.4}$ \\
2.7 & 9.25 & 8268.9 & $\sim\frac{1}{10^5}$ \\
4.2 & 7.75 & 495757.0 & $\sim\frac{1}{10^8}$ \\
6.1 & 5.85 & $\sim 10^8$ & $\sim\frac{1}{10^{12}}$ \\ \hline
\end{tabular}
\end{center}
\caption{Various mass splitting ratios $R$ for different values of $k$ and $c$.
$\frac{R_y}{r_z}$ gives the hierarchy between the two moduli.}
\label{T:1}
\end{table}
Table 1 has been constructed for different choices for the two moduli in a six dimensional model where the Planck to
TeV scale warping is achieved for every choice.
Such a hierarchical pattern of masses can simultaneously be obtained by including larger number of extra dimensions and
adjusting the different moduli according to Table 1. It has been explained in section 2 that for seven or higher dimensional
models the stack of 3-branes have energy scales clustered around TeV scale and Planck scale.

Here, using Table 1, we give few examples of how the fermion mass hierarchy can be addressed in multiply braneworld
models. Without loss of generality we can localize top quark to the TeV-brane at $(y,z)=(\pi,0)$, which
has a mass of $\approx$172 GeV. Then the fermion mass at the other TeV-brane, i.e. at $(y,z)=(\pi,\pi)$,
depends on the values of $k$ and $c$. For $k=0.7$ and $c=11.24$, the fermion field localized at $(y,z)=(\pi,\pi)$
would have a mass of $\sim$10 GeV, which is a reasonable value to explain the bottom quark mass of $\approx$4.2 GeV.
On the other hand, if we set $k=1.2$ and $c=10.75$, the fermion mass on the 3-brane at $(y,z)=(\pi,\pi)$
would be suppressed to $\sim$1 GeV, which is in the right ball park region to explain either the charm quark mass
($\approx$1.3 GeV) or the tau lepton mass ($\approx$1.77 GeV). Likewise, if we set $k=6.1$ and $c=5.85$,
the suppression on the 3-brane at $(y,z)=(\pi,\pi)$ is such that we can even explain the neutrino mass scale of 1 eV.
However, the values in Table 1 are generated for the case of six dimensional warped model, where we could only
explain the masses of two fermion fields. By generalizing this to seven dimensions, we will have four 3-branes
at around TeV scale and we also get additional freedom in the choice of the moduli $R_w$. Hence, in the seven dimensional
warped model, adjusting the value of the other modulus according to Table 1 we can explain four fermion masses. These
arguments can now easily be generalized in a braneworld model having more than seven dimensions where SM fermions of different masses
are localized on different 3-branes of the stack of clustered 3-branes around Tev scale through the usual localization mechanism \cite{koley}. 
The hierarchy in standard model fermion masses thus can be explained. 

\section{Conclusions}
In this work we have proposed a possible origin of the fermion mass hierarchy in the standard model through multiple warped geometry.
The requirement of the resolution of gauge hierarchy problem restricts the energy scale of half of the 3-branes in such model
around TeV while the remaining half are close to Planck scale. Using the mass generation mechanism of the standard model fermions 
proposed in \cite{GN},
where the right handed fermions are allowed to propagate in the bulk and both the Higgs boson and left-handed fermion fields of 
SM are confined to the 3-branes, we have shown that  we can adjust and tune various moduli to have different mass splittings once one mass parameter 
is determined by fixing the value of the only unknown parameter $Y_{6D}$. Assuming that the standard model fermions are 
localized at different 3-branes closely clustered around TeV
the slight difference in warping results into different contributions from the right handed fermion wave function 
in the Yukawa coupling which in turn leads to different masses of the SM fermion. The hierarchical mass pattern of the 
standard model fermions thus can be generated  by this mechanism. 
It is interesting to note from Table 1 that larger is the ratio between two moduli 
more is the splitting of fermion masses which are localized in those two branes.
The multiple warped model thus proposes a geometric understanding of the splitting of the masses of standard model fermions.

\end{document}